# Exponential and power law distribution of mass clusters in a (magnetic-like) deposition model of elongated grains in 2D piles

K. Trojan,[1,2] M. Ausloos,[1] and R. Cloots[3]

[1]SUPRATECS[*], Institute of Physics, B5, University of Liège,

B-4000 Liège, Euroland

[2]Institute of Theoretical Physics, University of Wrocław,

pl. M. Borna 9, 50-204 Wrocław, Poland

[3]LCIS[†], Institute of Chemistry, B6, University of Liège,

B-4000 Liège, Euroland

## Abstract

A generalized so called magnetically controlled ballistic rain-like deposition (MBD) model of granular piles has been numerically investigated in 2D. The grains are taken to be elongated disks whence characterized by a two-state scalar degree of freedom, called "nip", their interaction being described through a Hamiltonian. Results are discussed in order to search for the effect of nip flip (or grain rotation from vertical to horizontal and conversely) probability in building a granular pile. The characteristics of creation of + (or −) nip's clusters and clusters of holes (missing nips) are analyzed. Two different cluster-mass regimes have been identified, through the cluster-mass distribution function which can be exponential or have a power law form depending on whether the nip flip (or grain rotation) probability is large or small. Analytical forms of the exponent are empirically found in terms of the Hamiltonian parameters.

---

[*] SUPRATECS = Services Universitaires Pour la Recherche et les Applications Technologiques de matériaux Electro-céramiques, Composites, Supraconducteurs

[†] LCIS = Laboratoire de Chimie Inorganique Structurale, a member of the SUPRATECS Centre

# I. INTRODUCTION

Packed and disordered phases of two-dimensional (2D) hard disks is an old statistical problem [1–6] pertaining to fundamental questions on geometric (or entropic) phase transitions, e.g. packing and jamming [7, 8] appear in colloidal systems [9–11], vehicular traffic [12–16], and spin glasses [17]. Extensions toward binary, polydispersed or more complex objects have not received much attention [18–25]. Yet recent studies concerning granular piles suggest further investigations [26].

In constructing and describing granular piles it is crucial to consider that basic entities are not made of symmetrical units; if they can be thought to be made of hard cores [7], rice grains [27, 28] or sand grains [29, 30] can be hardly assumed to be spherical or disk shaped. Moreover intrinsic physical properties, like the surface charge, grain wetness, shape, .. lead to specific angles of repose [31, 32], jams [16], and arches [33–36]. These are seen in natural (sand, boulders,... piles), industrial (cement, coal, ... piles), food (wheat, carrot, rice, corn flakes, ... piles) or pharmaceutical (pill piles) cases. Beside macroscopic features other questions pertain to the inner *structure* of piles, like grain clusters, – not discounting physical effects about force distributions and the likes.

One unification basis for such investigations can be found in percolation theory initiated in the early forties [37, 38]. The ideas have been intensively developed and applied thereafter to many problems having found some equivalence in different subjects of science. The physics of granular matter is not an exception and has also drawn a great deal of information from phase transition theories through analytical work and simulation. Interestingly analogies and interpretations of percolation models in the field of granular media explain some of the basic questions on granular structures though many are still left without any answer.

In order to describe granular pile properties it is best to introduce degrees of freedom for the grain. The coupling of these degrees of freedom to an external field having the appropriate group symmetry allows to model various cases found in nature or laboratories. In physics, most degrees of freedom can be mapped into quantities, like "nips"(*nip* is the polish acronym for *Numer Identyfikacji Podatkowej* or Identification Tax Number). Such a "nip" main role is to break the spatial isotropic symmetry and to broaden the possible types of interactions between entities. The direction of a nip can e.g. represent an anisotropic feature, or the position of a grain with respect to neighboring entities, as well as a rotation process, or a dipolar effect, ... Although this is a very crude representation of a grain property, generalizations to more complex nip models are immediately imagined as it has already been done in describing granular segregation [39–42], decompaction [43], avalanches [44] through "spin-glass" phenomenology [45–49]. There are of course many other reports on how to build a granular pile [50–52].

Further analysis of granular effects in packings reveal indeed many connections with spin glass models, Ising models and percolation theory. For example the exchange energy $J$ describing a "spin-spin interaction" is analogous in granular matter to the contact energy between grains. For flows, a similar interpretation of $J$ can be found in Pandey et al. [53]. A constrained Ising spin chain has also been recently considered and studied as a toy model for granular compaction [54]. Extensions toward magnetic lattice gases [55], Blume-Emery-Griffiths models [56], higher spin and vertex models... can be easily envisaged.

In such a line, cluster percolation theory can be applied toward describing and later explaining several features or phenomena. Here below, the MBD (Magnetically controlled ballistic deposition) model [48, 49] is an extension of a well known ballistic deposition model recalled in [57–60] on which an extra degree of freedom is introduced.

The present granular model consists of elongated entities, characterized by one parameter, called "nip", indicating the longest grain direction. The nip can take two values $+1$ or $-1$ whether the grain long direction is vertical or horizontal. In some sense, the nip is like an Ising spin. A 2D pile of these entities is built by vertically depositing grains from an extended horizontal source, with the help of a Metropolis rule. This ballistic deposition process should not be confused with the binary Eden growth model on a 1D substrate [61], later on extended toward the case of a central seed growth [62] in 2D, nor with other chemical deposition models. During the present deposition process the grain can change its direction (the grain can rotate from vertical to horizontal, or conversely, whence the nip changes from $+1$ to $-1$ or conversely) with some probability $q$. The interaction between nips as well as the influence of an external field coupled to the nip were taken as those of the Ising Hamiltonian in [48] where $q = 1/2$ was imposed.

In this paper, the role of $q \neq 1/2$ is investigated first on the resulting pile structure, i.e. on the creation of clusters with a given nip sign, the pile density, the pile "magnetization" (to be defined), the fractal dimension, and the "percolation transition". In Sect. II, the algorithm rules are established and briefly commented upon. In Sect. III typical pile structures and numerical results for the density and the magnetization of the pile are presented. The fractal dimension and the cluster size distributions are discussed in Sect. III A-III D. Cluster mass regime distributions are identified, – the latter can have an exponential or power law form depending on whether the nip flip (i.e. grain rotation) probability is large or small. Analytical forms of the exponent are empirically found in terms of the Hamiltonian parameter and $q$. Finally, in Sect.IV, a brief discussion and a conclusion can be found.

## II. EXPERIMENTAL PROCEDURE

The algorithm creates a pile under a fixed probability $q$ for grain rotation, or nip change of sign, during the ballistic grain fall; the probability to choose the "up" direction (or $+1$ value) is $q$. The algorithm for arbitrary $q$ (so called $q$-MBD model, in contrast to the $\frac{1}{2}$-MBD model [48]) goes as follows:

1. a 1D substrate is built of nips (or hard elongated grains) with random direction chosen with probability $q$; periodic boundary conditions are used; although a grain is elongated, its size is such that only one grain occupies one lattice site at a time whatever the grain orientation

2. the underlying lattice is 2D triangular with vertical lines

3. when a site frontier of the cluster is reached by a falling grain the local energy gain

$$\beta E = -\beta J \sum_{<i,j>} n_i n_j$$

   is calculated, where $\beta$ is somewhat an irrelevant notation, since there is no temperature effect, and $n_j$ can take the value $+1$ or $-1$ depending on whether the grain long direction is vertical or horizontal (see Fig. 1).

4. If the "energy gain" $\Delta E$ is negative the grain sticks to the cluster immediately in its "nip" state. In the opposite case ($\Delta E > 0$) the grain sticks to the cluster with a probability $\exp(-\Delta \beta E)$ where $\Delta \beta E$ is the local gain in the energy. If the grain does not stick to the cluster it continues to fall down. Of course if the site just below the grain is occupied the grain immediately stops and sticks to the cluster.

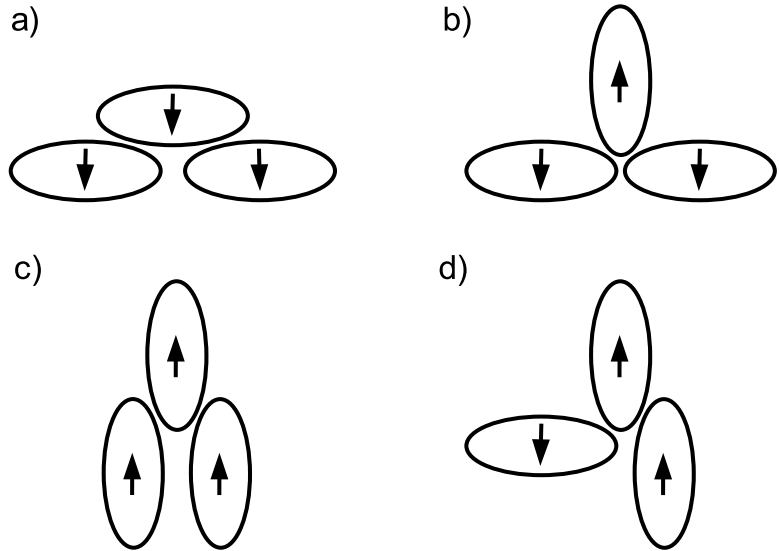

FIG. 1: Four typical deposition conditions, among many others, indicating, the relevance of the falling grain orientation with respect to the already stuck in the pile respective grain orientations (characterized by their "nip", indicated by an arrow) in order to calculate the local change in energy.

The exchange integral $J$ describing the nip-nip interaction can be thought in granular matter to be like some contact energy due to the peculiar surface roughness or some misorientation. It is expected that the orientation of the sticking grains can be inducing various clusters and constraints on the pile construction (Fig.1).

The above granular interpretation has surely some close similarity to classical spin depositions, like the magnetic DLA model [63, 64] and other magnetic deposition models. Whence by analogy with the magnetic case, we can consider that $J$ can be positive or negative, favoring during the pile construction, either a "ferromagnetic" or "antiferromagnetic" regime, i.e. a preferred piling through similar neighboring grains, or not, – similar in nip value, or in other words in grain orientation.

### III. NUMERICAL RESULTS

Results are all reported for a lattice of horizontal size $L = 100$ made of triangular cells having a vertical edge along which a grain falls down. When the pile height has reached a 100 lattice unit height, the algorithm is stopped. Every reported data point corresponds to an average over 1000 simulations.

#### A. Typical clusters

In Fig.2, 9 clusters having a $50 \times 50$ cell size taken from a pile grown under 9 different conditions are displayed. The central cluster ($q = 0.5; J = 0$) appears to have the smallest density.

The central-top ($q = 0.95$ and $\beta J = 0$) and central-bottom ($q = 0.1$ and $\beta J = 0$) clusters appear to have the same (low) density and only differ due to the fraction of nip with a particular direction. The right-top ($q = 0.95$ and $\beta J = 5$) and right-bottom ($q = 0.10$ and $\beta J = 5$) clusters correspond to the ferromagnetic-like case, where the interaction between nips induces the grains to prefer neighbors with the same nip value. The clusters appear to be tree-like. On the other hand, in the antiferromagnetic-like case, very compact and elongated structures appear to be produced: see the left-top ($q = 0.95$, $\beta J = -5$) and left-bottom ($q = 0.10$, $\beta J = -5$) clusters. Notice that most of the holes in the antiferromagnetic-like clusters are result of grain sticking when the nip belongs to the minority type for the whole pile. This lack of branching in the antiferromangetic case can be *a posteriori* understood.

#### B. Density

The overall density of a pile, as illustrated in Fig.3, is easily calculated from

$$\rho = \frac{\text{number of nips in the pile}}{\text{number of sites on the lattice}}, \tag{1}$$

in which obviously the number of lattice sites in the denominator $= 10000$.

Again distinguish differences between the antiferromagnetic-like (AF) $\beta J < 0$ and ferromagnetic-like (F) case $\beta J > 0$. Notice that a rather more loosely packed pile can be generated for $\beta J > 0$ (as in [48]): the density varies is greater than 0.38 but saturates at 0.47 when $q = 0.5$. The density can reach full compactness in the AF case (for $\beta J < 0$);

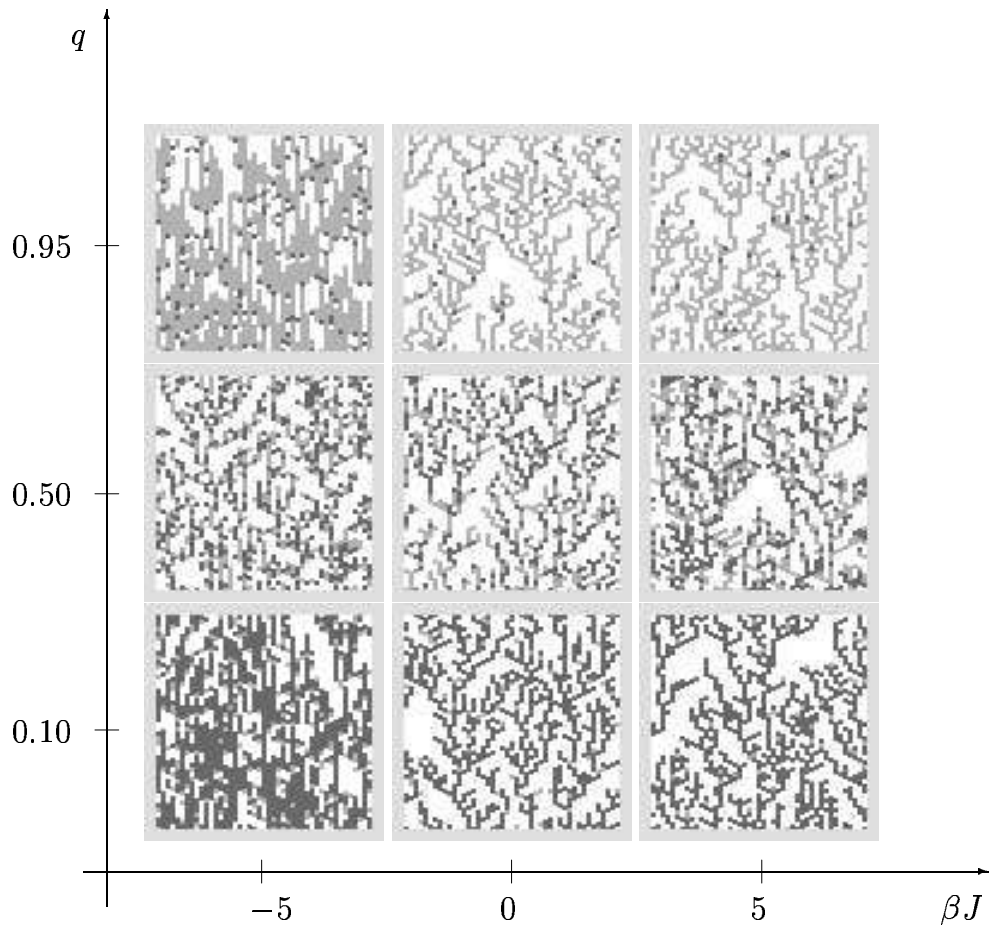

FIG. 2: Examples of clusters for several values $\beta J$ and $q$. Dark gray and light gray pixels indicate $-1$ and $+1$ nip values respectively.

its minimum ($\simeq 0.38$, which corresponds to the previously mentioned maximum density for the so called F-case) occurs at $q = 0.5$ for $\beta J = 0$.

The $+1$-nip $\rho_+$ density dependence on $\beta J$ and $q$ is presented on Fig.4. The density increases smoothly with $q$ and is quasi independent on $\beta J$. However, for $q \approx 1$, a (to be expected) relevant difference is seen between the AF and F cases. The $\rho_-$ dependence is easily deducted from Fig.4 by symmetry arguments.

Finally and for contrasting the above extremal density values with other simple models let us recall that monodisperse hard disks exhibit a freezing transition when the liquid volume fraction $(\phi) = 50\%$, and the solid about $55\%$. On the other hand a minimum density, equal to ca. 0.37, can be predicted by evaluating the probability of sticking [48]. The maximum density is close 0.933, i.e. the hard disk melting density [1]. The presence of a minimum (or maximum) density can be understood for a finite negative (or positive) $\beta J$ value to occur at $q$=0.5.

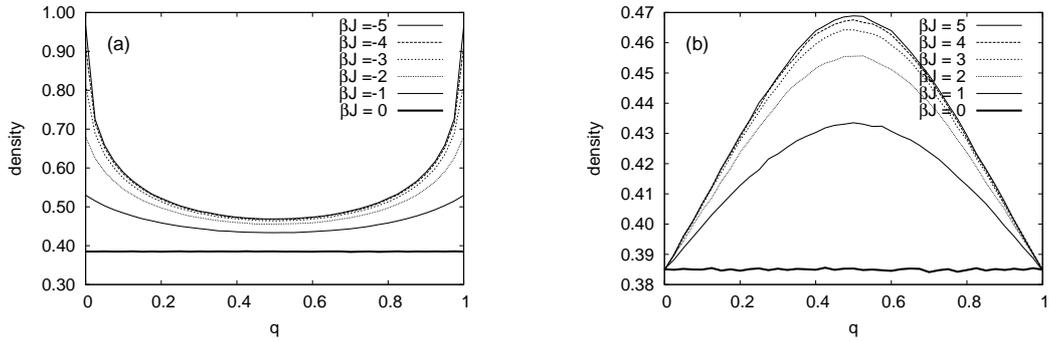

FIG. 3: Dependence of the pile density on $q$ for different $\beta J$ values, in the so called (a) ferromagnetic-like case, (b) anti-ferromagnetic-like case.

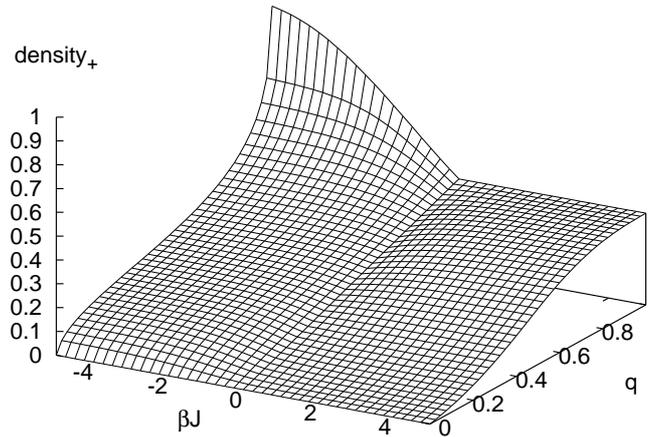

FIG. 4: Dependence of the $+1$ nip pile density $\rho_+$ on $\beta J$ and $q$ parameters.

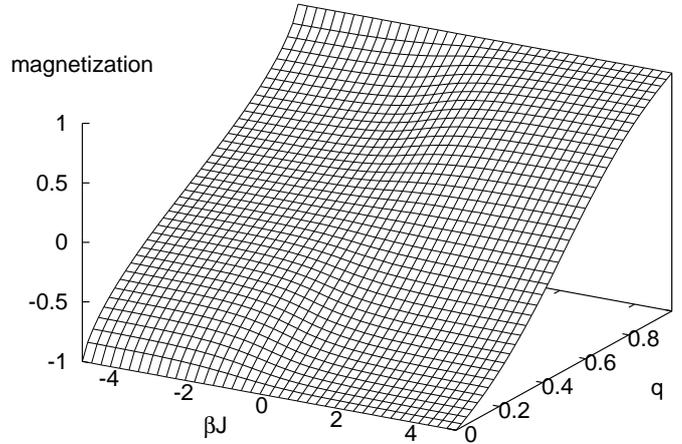

FIG. 5: Dependence of the so called magnetization on $\beta J$ and $q$.

## C. Orientational order parameter

In magnetic lattice gases [55, 56] there are two order parameters: the density and the magnetization. As the magnetization can be defined for such systems, we consider the difference between orientation types in the pile through

$$M = \frac{\rho_+ - \rho_-}{\rho_+ + \rho_-} \qquad (2)$$

as shown in Fig.5 as a function of $\beta J$ and $q$.

One can note a very smooth dependence of $M$ on $q$ with only small departures from linearity within a large range of $q$, due to the changes in the interaction strength. There is a marked symmetry between the regions $(\beta J > 0, q < 0.5)$ and $(\beta J < 0, q > 0.5)$.

## D. Fractal dimension

In order to compare the standard MBD model with the $q$-MBD model we have checked the fractal dimension [65, 66] of the piles in different parameter regions through the (box counting) technique [65, 66]. This consists in covering, without overlapping, the whole cluster by squares of same sizes, and counting the number of squares which have at least one nip $+1$ (or $-1$), then modify the square sizes. We have distinguished between the fractal dimension of the global pile and that of clusters made of $+1$ only and $-1$ only nips. From the best linear fit to the data, i.e. log(number of squares with a nip) vs. $-\log$(square size), the fractal dimension is obtained through the slope. The asymptotic fractal dimension for the whole cluster, i.e not distinguishing between the nip signs, varies between $1.95 \pm 0.04$ and $2.02 \pm 0.02$, hence is equal to 2 whatever $\beta J$, taking into consideration the error bars. This value is similar to that obtained in binary Eden growth models [61, 62] and in simple ballistic deposition models [60].

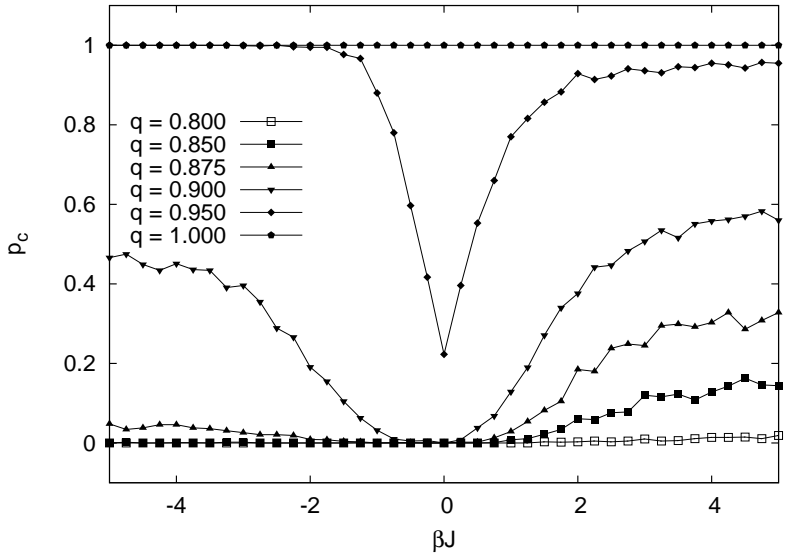

FIG. 6: Dependence of the percolation probability $p_c$ on $\beta J$ for several values of the $q$ parameter.

### E. Percolation

In this section we present results concerning the percolating cluster (a percolating cluster is defined as the cluster extending from the bottom to the top of the pile without any disconnection) of a typical entity, e.g. the set of $+1$-nips. In order to do so every pile built by the algorithm has been checked in order to determine whether there is a percolating cluster of $+1$ nips. At fixed $q$ and $\beta J$ the fraction $p$ of piles consisting of a given type percolating cluster was computed. The value at wich $p$ vanished from above is thought to be representative of the critical percolation $p_c$ for such a cluster type.

The behaviour of $p_c$ with respect to $\beta J$ for several values of the $q$ parameter is shown in Fig.6. Observe that there is no percolation for $q < 0.75$ in the F-case and 0.85 in the AF-case. The former case is very similar to that observed in standard percolation theory (i.e. without interaction between entities). A bird eye view of the dependence of the percolation probability on $\beta J$ and $q$ is shown in Fig.7. Notice a 'hollow' near $\beta J = 0$, the case without interactions; it has the highest $p_c$ value. We stress the asymmetry of the percolation probability as a function of $\beta J$.

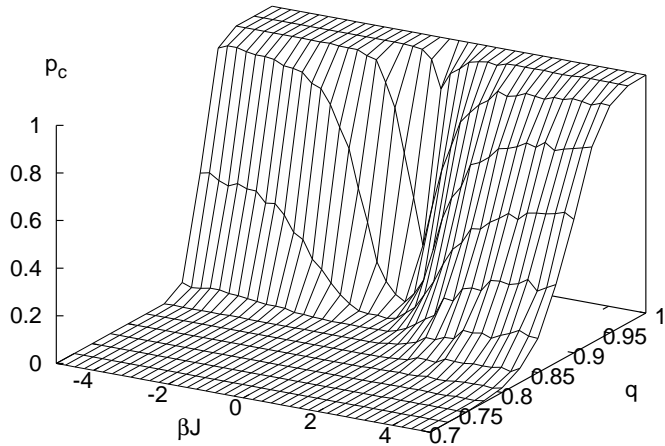

FIG. 7: Bird eye view of the dependence of the percolation probability on $\beta J$ and $q$.

The differences between AF and F are understood if one recalls that the density in the AF case is much more sensitive to a change in $q$ than in the F case, for which a change of $q$ induces only a very small variation of the density, – recall that the density in this case is $[0.38; 0.47]$ (Fig.3). On the other hand a similar variation in the $q$ parameter generates piles with a wide interval of possible densities in the AF case (see Sect. III B).

### F. The mass of clusters

The number $N_s$ of clusters with mass s (called "size" here below) is shown in Fig.8 for low $q$ values. We may postulate

$$N_s(s) \propto e^{-k_E s} \qquad \text{for small} \qquad q, \tag{3}$$

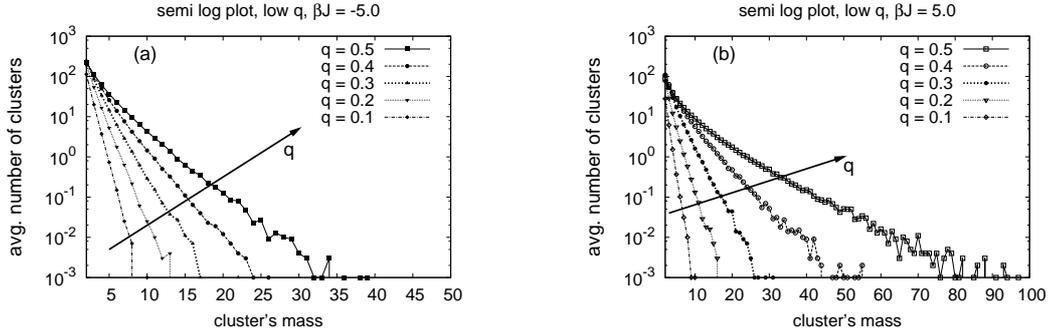

FIG. 8: Examples of the $N_s(s)$ dependence for low values of the $q$ parameter presented on semi-log plots: (a) $\beta J = -5$; (b) $\beta J = 5$.

where $k_E$ is a constant for a fixed $(\beta J, q)$ pair and $s$ is the cluster size. Observe that the slopes of $N_s$ for fixed $q$ in the AF case are higher than in the F case, indicating that the existence of larger clusters can be expected in the F more than in the AF case. On Fig.8a one can also notice that for high values of $q$ (see e.g. $q = 0.5$) $N_s$ might not be a pure exponential law at low $s$.

Further analysis (Fig.9) of $k_E$ reveals that its behaviour can be represented as

$$k_E = 4\exp(-A_E q), \quad (4)$$

as long as $q$ is less than $p_c$, i.e. 0.75 or 0.85 depending on the $\beta J$ sign.

Studying the slopes in the interval $q \in [0:0.6]$, it appears that $A_E$ is a simple function of $\beta J$ (Fig.10): $A_E$ is constant in the AF case but is growing toward a value $\approx 6.5$ in the F region, according to

$$A_E = 4\theta(-\beta J) + F(\beta J)\theta(\beta J) \quad (5)$$

where $\theta(x)$ is the Heaviside step function and an excellent $F(\beta J)$:

$$F(\beta J) = 4.02 + 2.5\tanh(0.52\beta J). \quad (6)$$

Summarizing we have for the ERL region:

$$N_s(s) \propto e^{-4\exp(-A_E q)s}, \quad (7)$$

where $A_E$ has been presented in Fig.10. Such a double exponential law with a (relatively simple) strength dependent amplitude has still to be understood.

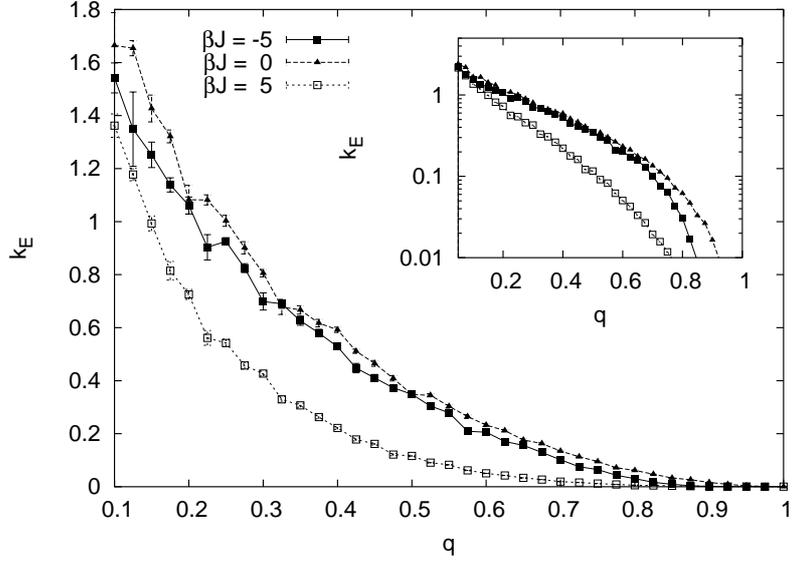

FIG. 9: The parameter $k_E$ in the $N_s$ exponential law dependence. The inset shows the same data on a log-log plot.

For high $q$ values the $N_s$ dependence (Fig.11) stops to be an exponential law and becomes a power law. Let us postulate

$$N_s(s) \propto s^{-k_P} \qquad \text{for} \qquad q \approx 1. \tag{8}$$

This postulate holds down to $q = 0.85$ and $q = 0.75$ respectively for the AF and F cases. Moreover the slopes on a log-log plot appear to be close to $-2.0$ and $-1.5$ for the AF and F cases respectively.

Fig.12 exhibits in a 3D way display, the dependence of the power law exponent $k_P$ on the interaction strength $\beta J$ and $q$. It has a less trivial dependence than $k_E$. Observe markedly large data scattering (or surface roughness) in $k_P$ value fluctuations for $q < ca.$ 0.6. Yet the error on $k_P(\beta J, q)$ depends on the $q$ parameter, but is quasi $\beta J$ independent (Fig.13). One can note that the error on $k_P$ is decreasing with increasing $q$ but is low enough at $q > 0.6$ to claim that there is a pure power law regime above a critical $q$, being $q_c = 0.85$ and $q_c = 0.75$ respectively for the AF and F cases. The $k_P$ behavior of the AF case has indeed an inflexion point for such $q$ values.

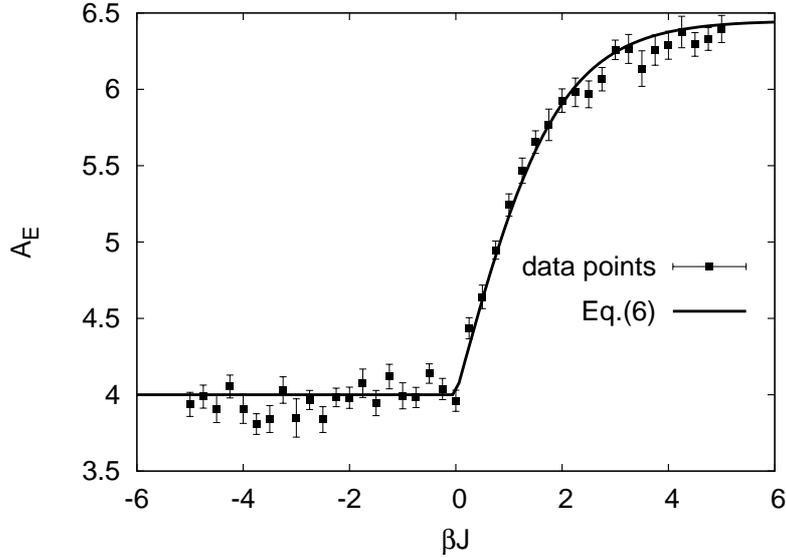

FIG. 10: The dependence $A_E$ on $\beta J$, with an empirical fit as described in the text.

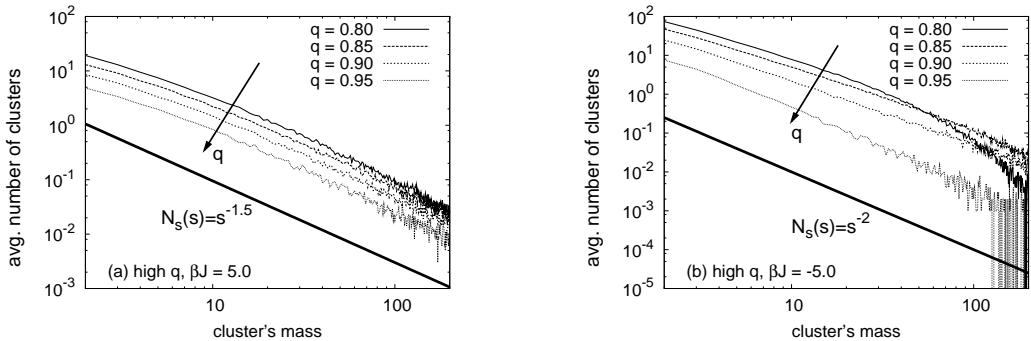

FIG. 11: Examples of the $N_s(s)$ dependences for high values of the $q$ parameter presented on log-log plots; (a) for an antiferromagnetic-like case, i.e. $\beta J = -5$; (b) for a ferromagnetic-like case, i.e. $\beta J = 5$. Characteristics slopes ($-1.5$ and $-2.0$) are indicated to show that such cases are different indeed.

### G. The lacunarity of clusters

Analogously to $N_s$ we can define $N_h$ as the number of hole clusters (or lacunes) [67–69] (see Fig.14). Unlike for the nip clusters there is no difference in $N_h$ dependencies between low and high $q$ values, – all $N_h$ dependences seem to follow a *power law*, with a non trivial exponent $k_h$ as measured from the slope on such log-log plots. The dependence of $k_h$ is shown in Fig.15 as a function of the interaction strength and $q$. For very compact AF piles the exponent exceeds 2.6, but in F piles the exponent is equal to at most 2.1. One can recognize that the higher the pile density the higher the exponent $k_h$.

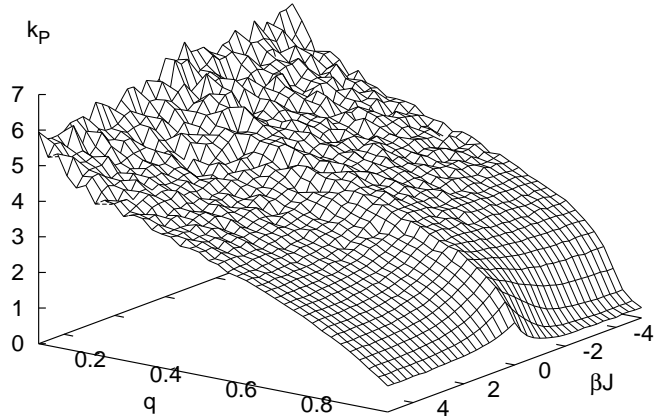

FIG. 12: Dependence of the power law exponent $k_P$ for cluster distributions as a function of $\beta J$ and $q$.

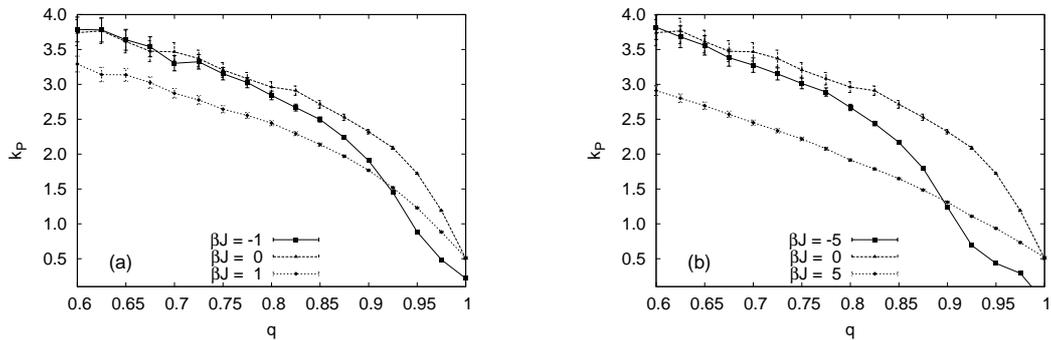

FIG. 13: The dependence of $k_P$ for (a) low and (b) high interactions, in the power law size (mass) distribution regime.

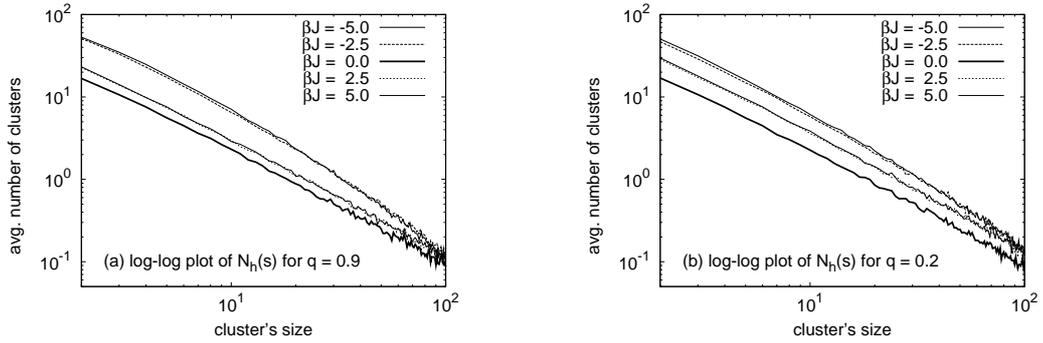

FIG. 14: The dependence of $N_h$ on a log-log plot for high (a) and low (b) values of $q$.

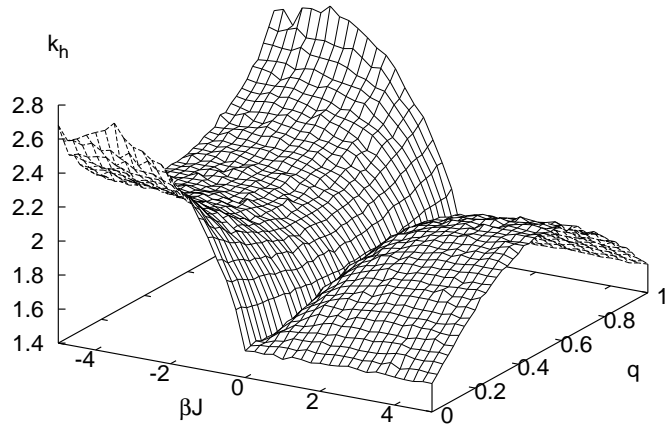

FIG. 15: The dependence of the power law exponent $k_h$ on $\beta J$ and $q$ for the distribution of hole clusters in a pile.

## IV. CONCLUSIONS

Even though we investigated finite size systems, thereby accepting finite error bars, it appears that some definite conclusions can be reached from this magnetically controlled ballistic rain-like deposition (MBD) model in order to mimic granular pile structures and basic properties in 2D. First from a statistical mechanics point of view, the model has similarities with various classical magnetic lattice gas, kinetically grown cluster and percolation models.

Yet it differs by specificities in order to approach granular piling in a simplified whence different way than the original spin glass (and Tetris) models. The grains, characterized by a two-state scalar degree of freedom, called "nip", similar to a spin as their interaction is concerned, can rotate during their fall. It is known that grain rotation is a major problem in explaining pile features. Results are here above presented in order to search for the effect of such grain rotation from vertical to horizontal and conversely assuming a fixed probability for doing so during a granular pile building. Two different cluster-mass regimes have been identified, through the cluster-mass distribution function which can be exponential or have a power law form depending on whether the nip flip (or grain rotation) probability is large or small. The surface growth exponents have not been studied [70, 71].

It is easily noticed that the higher the nip-nip interaction strength the bigger is the difference between the various piling cases. Whence it seems that one can distinguish pile growth conditions with respect to $q < q_c(\beta J)$ and $q > q_c(\beta J)$. Analytical forms of the exponent are empirically found in terms of the Hamiltonian parameter. Moreover such critical values are identified as for phase transition cases to depend in a non trivial way from the interaction strength. The analytical form needs further work, but the critical values 0.85 and 0.75 can received a simple interpretation, when realizing that $0.85 \simeq 0.83 \simeq 5/6$, while 0.75 of course $= 3/4$, indicating the most probable basic cluster to occur from a free energy like point of view on such a regular lattice. It remains to be seen if this can be proven experimentally.

We are aware that there are many criticisms which can be raised on the model as a so called simulation of grain piling : e.g. there is an obvious lack of elasticity [1], i.e. there is neither a back-scattering effect nor relaxation due to the kinetic energy of the falling grain and the stress redistribution.

The external pressure field coupled to the so called $C_x$ of the grain is also missing. Other electrostatic and magnetic interactions have been neglected. In contrast see the particle-cluster aggregation in presence of dipolar interactions by Pastor-Satorras and Rubi [72]. Yet the geometry is similar to that of silo structures [42]. We have not taken into account interaction with the silo wall since periodic boundary conditions are used. It is useful to point out that the wall could be taken as an external (magnetic or more generally) force field. See in this spirit studies of irreversible growth of a magnetic material confined between parallel walls where competing surface magnetic fields act [71]. The interplay between confinement and growth mode leads to a physically rich phase diagram (on the H-T plane) that exhibits a localization-delocalization transition in the interface. Such a question needs further investigation in our model when introducing an external field, non uniform in contrast to [48, 49].

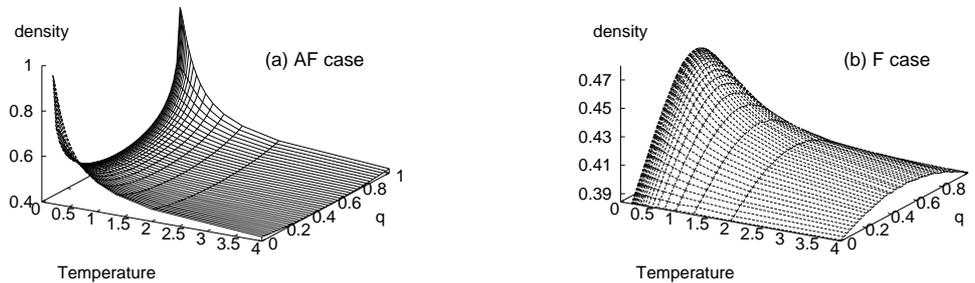

FIG. 16: The dependence of the pile density on temperature and $q$ parameter, – a thermodynamic interpretation.

One might also wonder whether the wetness of the grain cannot be taken into account through the $\beta$ "temperature factor". In so doing of course, all properties like the fractal dimension, density, "orientational order parameter", should be temperature dependent, and transitions might be found, as seen in magnetic Eden and/or diffusion-limited aggregation. It was found, e.g. in [72] from the orientation probability density of the dipole direction in the clusters that the ordered state found at low temperatures diminishes when the temperature increases, due to the disorder induced by the fractal geometry of the aggregates. The same was noticed in the magnetic binary Eden model [73–75] where the minimum in the fractal dimension corresponds to the onset of magnetization, even when there are finite size effects [76]

One interpretation of the $\beta J$ parameter is a thermodynamic one, i.e $1/\beta J = kT/J$. Assuming $J$ being fixed interactions we can imagine or interpret the behavior of the pile at different temperatures. For the high temperature approximation the AF and F cases do not much differ (Fig.16) – this refers to non-interacting-like spin cases. The lower the temperature the bigger are the differences in the density between these two states (AF and F). At low temperature and for marginal $q$ values the AF case is characterized by high density piles, – very compact ones. On the other hand the most compacted piles occur in the F case or for low temperatures for $q \approx 0.5$. A rather fast drop in the highest density (for fixed $\beta J$) is observed with increasing temperature, which seems a reasonable expectation, – i.e. the system behavior can be extrapolated as toward that of a granular gas case.

**Acknowledgments**

KT is partially supported through an Action de Recherches Concertée Program of the University of Liège (ARC 02/07-293).